\documentclass[preprint,12pt,unsrt,compress]{elsarticle}
\usepackage{graphicx}
\usepackage{a4wide}
\usepackage{epstopdf}
\usepackage{booktabs}
\usepackage{natbib} 
\usepackage{multirow}
\usepackage{amssymb}
 \usepackage{amsthm}
 \usepackage{amsmath}
 \usepackage{lscape}
 \usepackage{hyperref}

\journal{Elsevier}

\begin{document}

\begin{frontmatter}

\title{
Grandpa, grandpa, tell me the one about Bitcoin being a safe haven: Evidence from the COVID-19 pandemics
}

\author[ies,utia]{Ladislav Kristoufek} \ead{LK@fsv.cuni.cz}

\address[ies]{Institute of Economic Studies, Faculty of Social Sciences, Charles University, Prague, Czech Republic, EU}
\address[utia]{The Czech Academy of Sciences, Institute of Information Theory and Automation, Prague, Czech Republic, EU}

\begin{abstract}
Bitcoin being a safe haven asset is one of the traditional stories in the cryptocurrency community. However, during its existence and relevant presence, i.e. approximately since 2013, there has been no severe situation on the financial markets globally to prove or disprove this story until the COVID-19 pandemics. We study the quantile correlations of Bitcoin and two benchmarks -- S\&P500 and VIX -- and we make comparison with gold as the traditional safe haven asset. The Bitcoin safe haven story is shown and discussed to be unsubstantiated and far-fetched, while gold comes out as a clear winner in this contest.
\end{abstract}


\end{frontmatter}


\section{Introduction}

History of Bitcoin is tightly connected to its detachment and independence from the standard financial markets and the proclaimed properties that should make it serve as the `digital gold' \cite{Courtois2014}. An important implication of such status is Bitcoin potentially being a safe haven asset either in addition to or as a replacement of gold itself that has served as such for decades \cite{OConnor2015}. The safe haven asset is an asset that capital can take a refuge in when other assets are in distress. The distress situation of the other assets is a clear distinction from being a good diversifier, which traditionally leads towards a low or even negative correlation with other assets in the Markowitz logic of portfolio construction \cite{Markowitz1952}. An asset might be considered a safe haven if its correlation with other assets during the turbulent periods is lower (at least not higher) than during the calm periods \cite{Baur2010,Beckmann2015,Klein2017,Klein2018}.

\begin{figure}[!htbp]
\begin{center}
\begin{tabular}{cc}
  \includegraphics[width=0.45\textwidth]{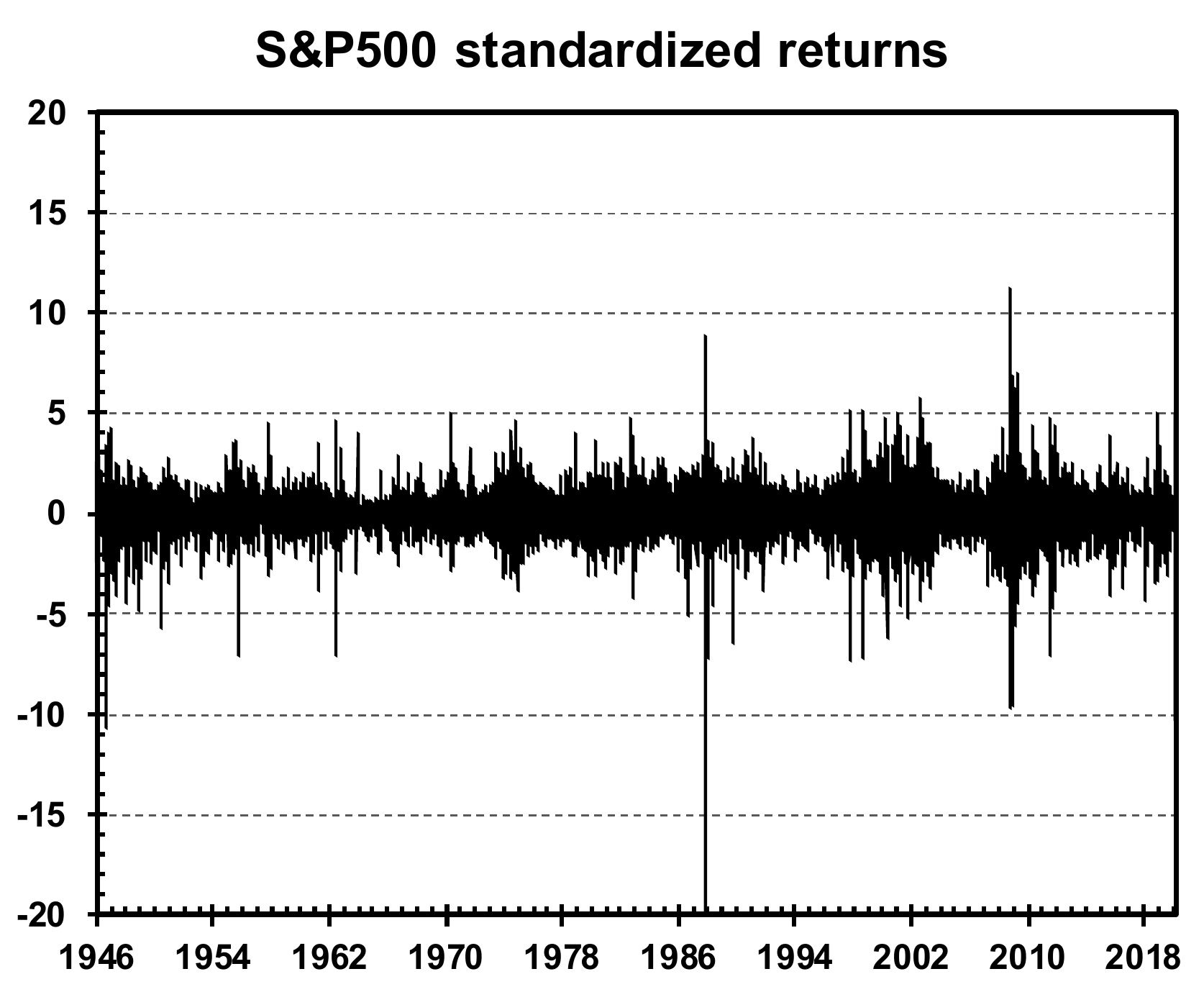}&\includegraphics[width=0.45\textwidth]{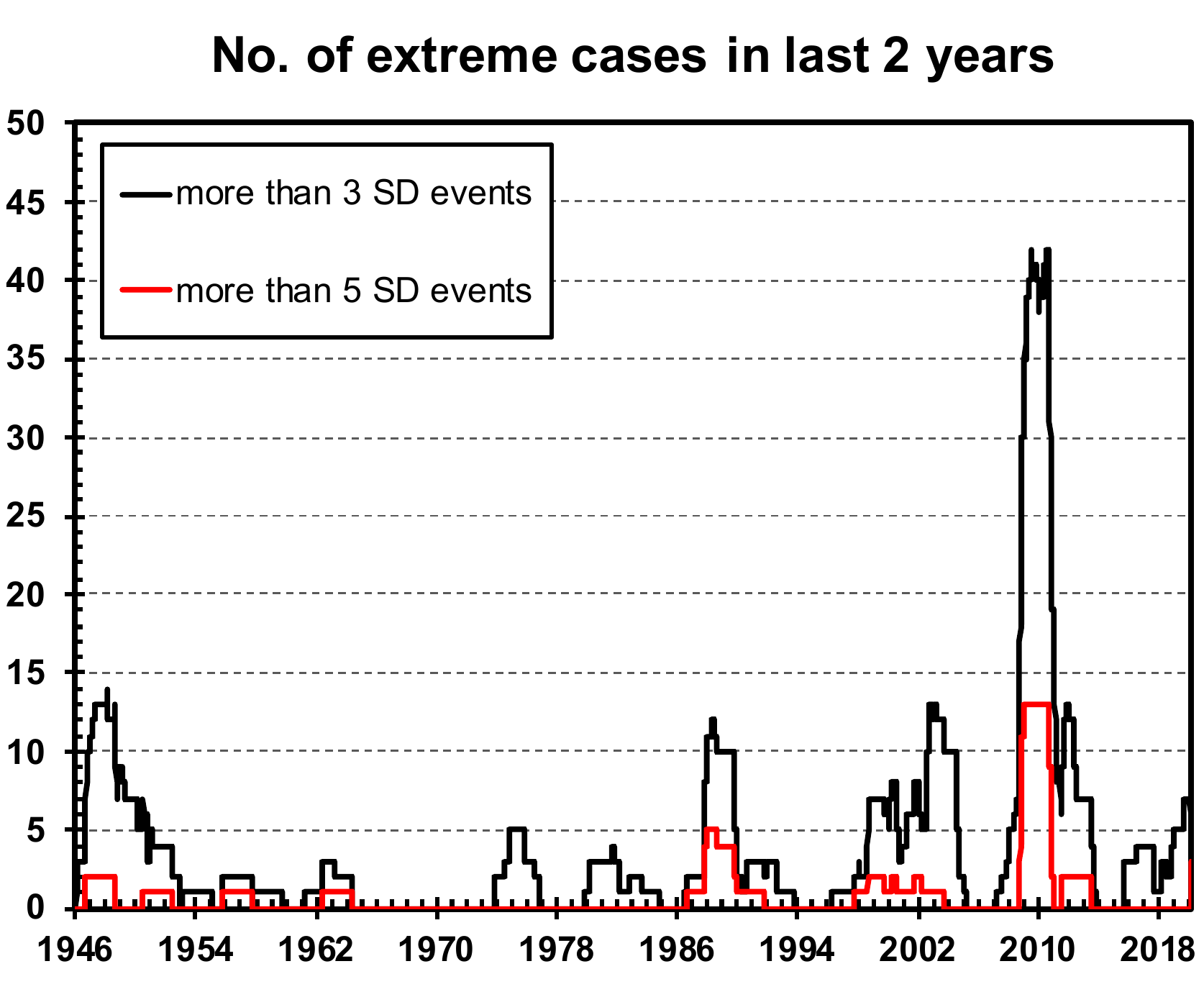}
\end{tabular}
  \caption{
Historical extreme events of S\&P500. (Left) Logarithmic returns of S\&P500 between 1 Jan 1946 and 12 March 2020 demeaned and standardized with historical mean and standard deviation over the whole examination period. (Right) Number of extreme returns over 3 or 5 standard deviations. Cumulative count on a rolling window of 500 days is shown. 
}
  \label{SP}
\end{center}
\end{figure}

The safe haven status of Bitcoin is one of its cornerstones and narratives in the financial part of the crypto-community and it has been a popular topic in the scientific literature as well \cite{Bouri2017,Wang2019,Smales2019,Selmi2019,Urquhart2019}. However, its validity had been, by definition, very difficult to properly discuss and test as empirical tests had lacked the essential part of the safe haven definition -- the financial markets in distress. As Bitcoin was developed in 2008 and 2009 \cite{Nakamoto2008}, and its first legendary pizza transaction took place in March 2010 and gained some larger public attention only by 2013, still mostly due to its controversial aspects (such as Mt. Gox, darknet, and Silk Road), it had avoided the most turbulent times of the global financial crisis. And it had taken until the middle of 2016 for the Bitcoin markets to reach a stable daily traded volume of more than \$100 million. To illustrate the historical perspectives, Fig. \ref{SP} shows the S\&P500 standardized daily logarithmic returns back to the beginning of 1946 where we find historical critical events with episodes of numerous negative returns of more than five historical standard deviations (the series is demeaned and standardized by histocal mean and standard deviation of the dataset between 1 Jan 1946 and 12 March 2020). To put the extreme events into a better perspective, the right panel of Fig. \ref{SP} shows a number of extreme events above three and five standard deviations on a sliding window of two trading years (500 trading days). There, we see that since 1987, there have been only few periods of time without these 5-SD critical events. Yet, one of these periods has been between 2013 and 2020, i.e. the period of Bitcoin's existence with some palpable trading volume and usage. It has been only the days of March 2020 that experienced severe losses of the financial markets due fear and uncertainty connected to the COVID-19 (coronavirus diseases 2019 of virus SARS-CoV-2) pandemics originating in China at the breaking of 2019 and 2020 and spreading rapidly and widely to other continents.

Even though the spread of the virus had been assumed to be possibly locally contained, its unprecedented spread has caused a widespread panic in the global society which quickly translated to sell-outs and havoc on the financial markets. Purely statistically (and perhaps cynically) speaking, this creates a unique opportunity to test the safe haven properties of Bitcoin and compare it with gold as the traditional safe haven of choice.

\section{Results}

We study the interconnection between Bitcoin (BTC) and two benchmarks -- the Standard \& Poor's 500 (S\&P500) index as a representative of the global financial markets and CBOE Volatility Index (VIX) as a measure of the markets uncertainty. We use publicly available data from finance.yahoo.com and with this respect, we also utilize the Bitcoin prices provided there (these reflect the CoinMarkepCap.com data) which restricts the analysis to start from 16 Sep 2014. The ending is 12 March 2020. As Bitcoin is traded on the 24/7 basis and stocks are not, we use the close-close logarithmic returns\footnote{It certainly is up to discussion whether to use returns for VIX as well. We have considered this possible issue and performed the analysis on both levels and returns of the VIX index. The results are qualitatively the same. Note that the distinction between logarithmic and original series plays no role here as we apply a quantile-based method (and logarithm is a monotonous transformation).} (rather than open-close) to include the weekend movements of Bitcoin. This gives us 1380 daily observations.

As the safe haven property is similar to being a diversifier, i.e. having a low correlation with other assets, but only during critical times, we approach it from a simple perspective of examining correlations between Bitcoin and the other two assets -- S\%P500 and VIX -- during critical events. We treat the critical events as rarely occurring, negative events, i.e. events in the (very) low quantiles of the distribution of the baseline asset. For this purpose, we utilize the quantile correlation as introduced by Li \textit{et al.} \cite{Li2015}. For statistical validity, we estimate the quantile correlation coefficient on 1000 bootstrapped samples (resampling the time index with a replacement) so that we can present not only a point estimate but also confidence intervals.

\begin{figure}[!htbp]
\begin{center}
\begin{tabular}{cc}
  \includegraphics[width=0.45\textwidth]{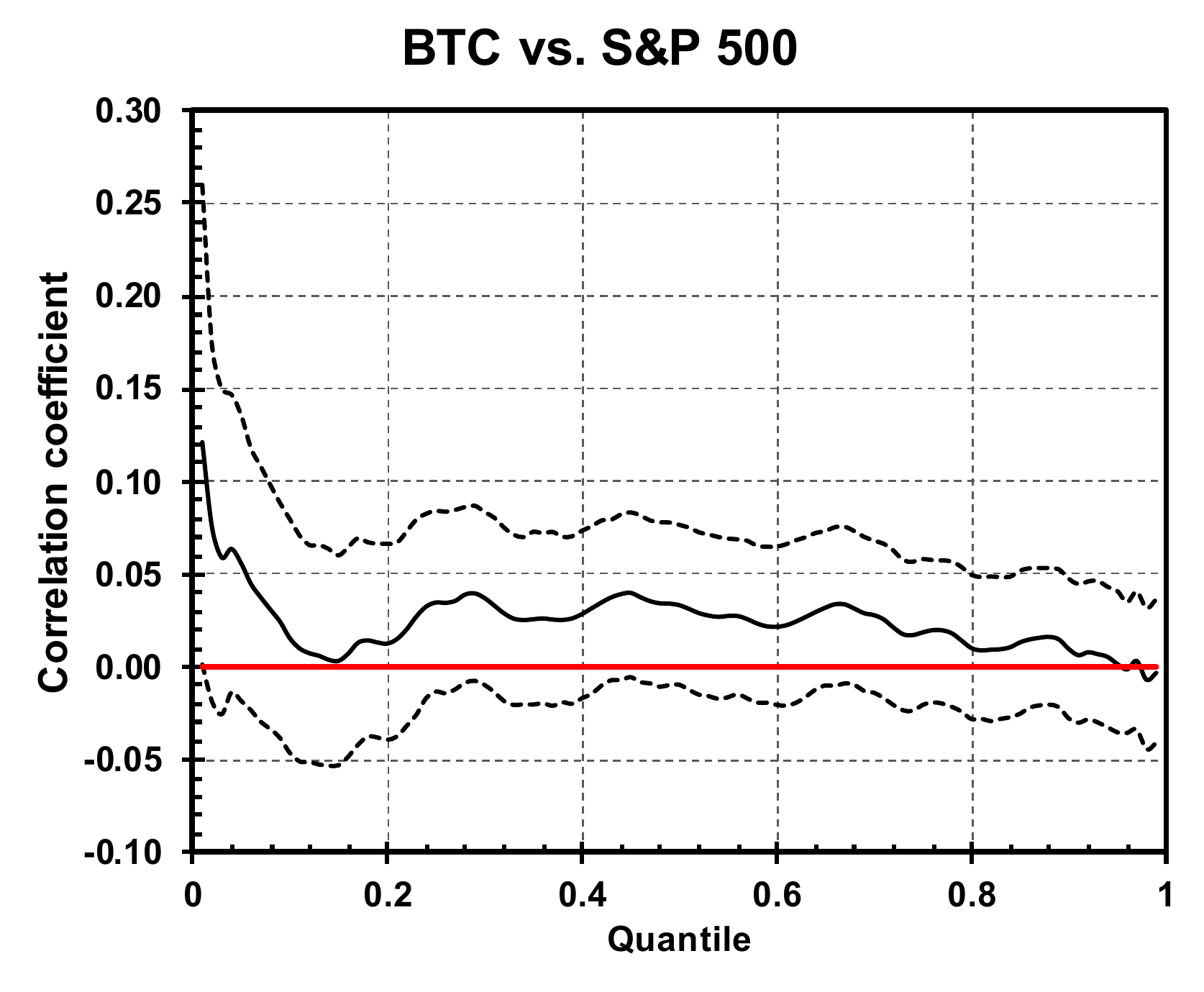}&\includegraphics[width=0.45\textwidth]{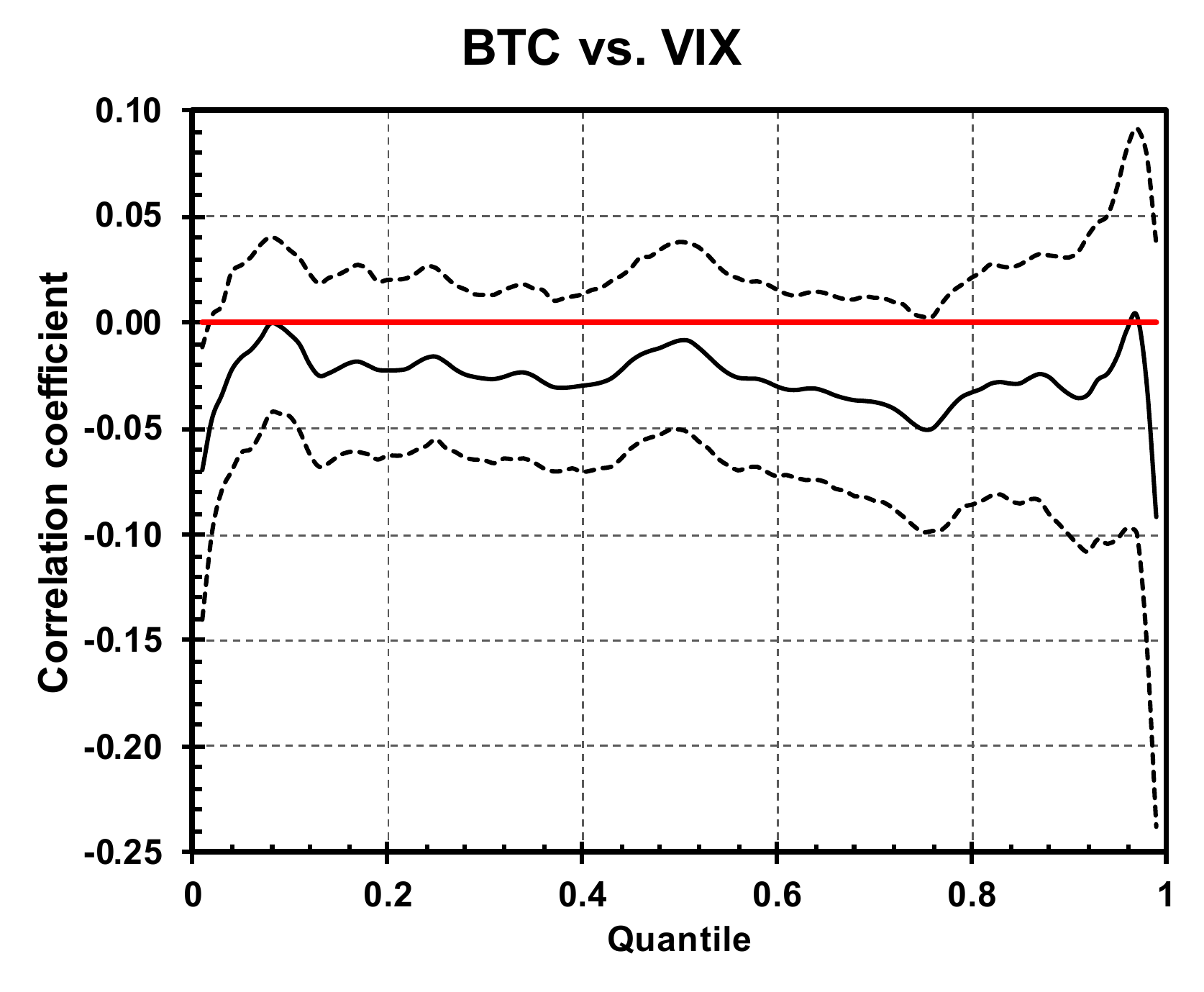}
\end{tabular}
  \caption{
Quantile correlations for Bitcoin. (Left) Quantile correlations between Bitcoin and the S\&P500 index. The quantiles on the $x$-axis are with respect to the S\&P500 index. The low quantiles show the extreme negative events. Black bold curve shows the mean value of 1000 bootstrapped estimates. The dashed curves show the 90 \% confidence intervals based on the bootstrapped estimates. (Right) Quantile correlations between Bitcoin and the VIX index. The quantiles on the $x$-axis are with respect to the VIX index. The high quantiles show the periods of high uncertainty. The other notation holds.
}
  \label{BTC}
\end{center}
\end{figure}

In Fig. \ref{BTC}, we see the quantile correlations between BTC and S\&P500 (left), and between BTC and VIX (right). The quantile here represents the conditional quantile of the latter asset in the pair, i.e. either S\&P500 or VIX. We find that BTC is a good diversifier with respect to S\&P500 in the calm and bullish times, i.e. in the bulk of the distribution and more generally between quantiles 0.2 and 1, with correlations very close to zero while the 90 \% confidence intervals include the zero correlation. For the very low quantiles below 0.1, the correlation increases up to more than 0.1. The combination of low quantiles of S\&P500 and  a positive correlation signals that BTC drops together with the stock market if the situation is critical. Note that the size of the correlation is still quite low but still well above the levels during the calmer periods. For VIX, which represents overall mood on the market and expected future uncertainty, we need to look at the high quantiles as it holds that the higher VIX is, the higher the uncertainty. For a safe haven asset, we would expect a low or positive correlation at least in these high quantiles, or ideally positive correlations for all quantiles. We observe a similar picture as for the S\&P500 case as the correlation is very close to zero for most situations but it drops markedly for the times of high uncertainty, which is not a sign of a safe haven.

\begin{figure}[!htbp]
\begin{center}
\begin{tabular}{cc}
  \includegraphics[width=0.45\textwidth]{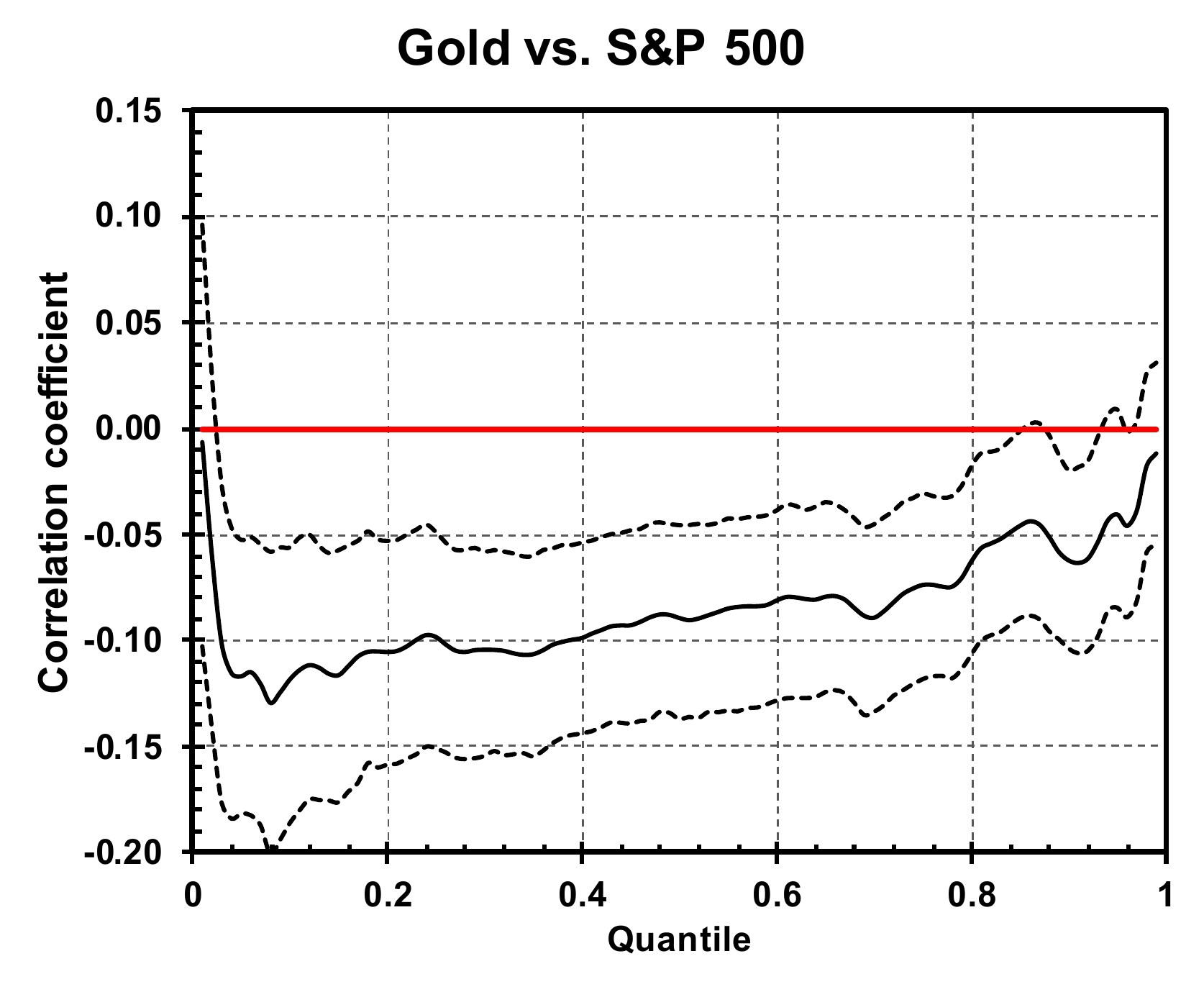}&\includegraphics[width=0.45\textwidth]{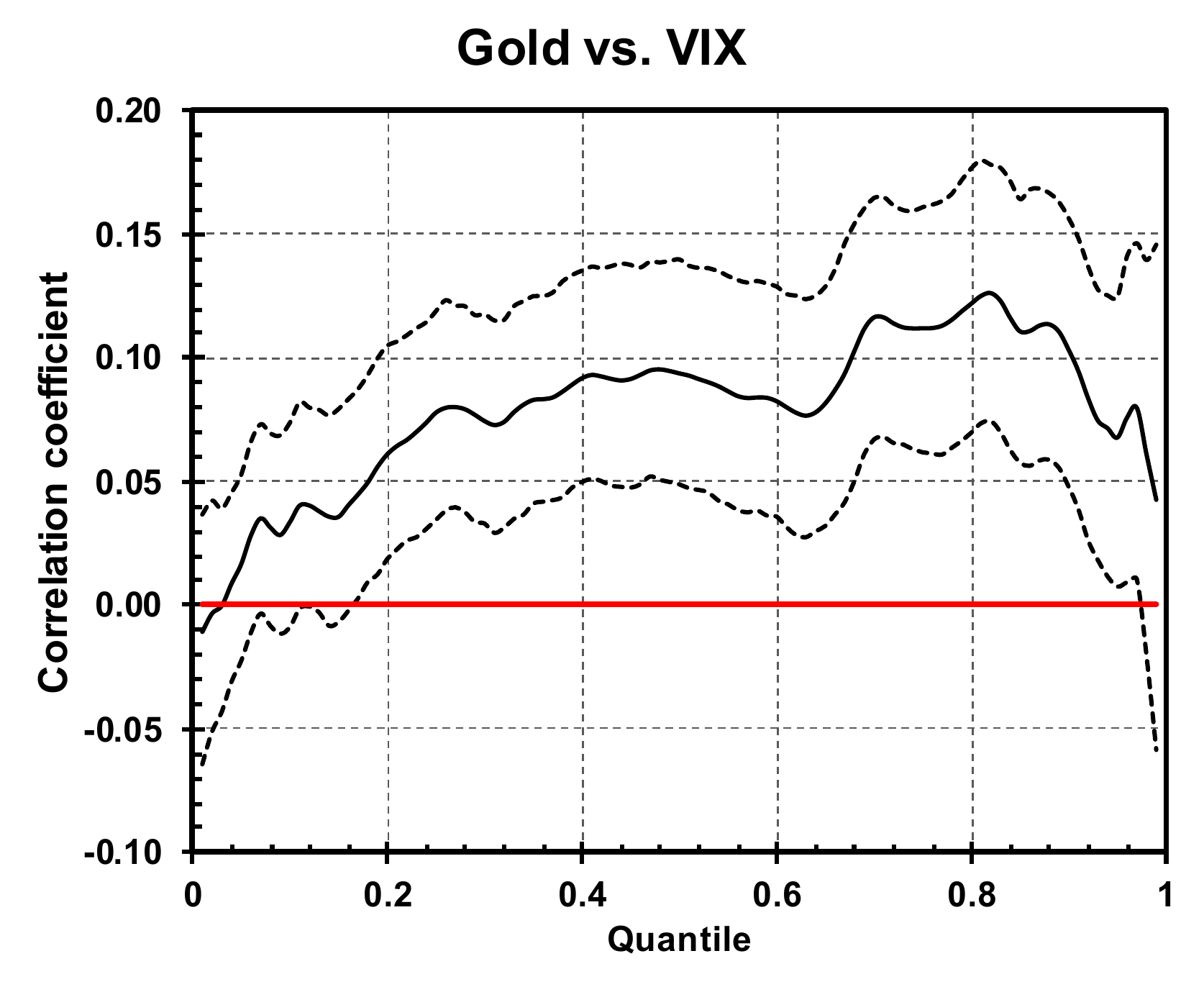}
\end{tabular}
  \caption{
Quantile correlations for gold. (Left) Quantile correlations between gold and the S\&P500 index. (Right) Quantile correlations between gold and the VIX index. The notation holds from Fig \ref{BTC}.
}
  \label{Gold}
\end{center}
\end{figure}

Comparing the results to the traditional safe haven of gold (Fig. \ref{Gold}), we see a bit different picture. In the bulk of the distribution, gold is negatively correlated with S\&P500 and even though its correlation increases during the extreme negative events, its estimate still remains below the zero correlation. With respect to VIX, gold is positively correlated with it in the distribution bulk and even though its correlation decreases for the most uncertain periods, it still remains above zero.

\section{Discussion and conclusions}

The COVID-19 epidemics is the first global economic and financial earthquake that took place during the existence and the actual use and wider knowledge of Bitcoin which made it possible to put the claims of Bitcoin being a safe have asset to an actual empirical examination. We study the quantile correlations between Bitcoin and a pair of global financial benchmarks -- the S\&P500 index as the stock market benchmark and the VIX index as a measure of uncertainty and future expectations. What we find is that Bitcoin can be easily considered as a good diversifier as its correlation with S\&P500 is close to zero for most of the quantiles. However, its correlation increases markedly during the turbulent periods of the S\&P500. The mirror result is observed for its relationship with the VIX index as the correlation remains close to zero for most quantiles again but drops for the most uncertain times. However, even the extreme-quantile correlations between Bitcoin and either S\&P500 or VIX still remain rather low (in absolute terms) and one needs a comparison to fairly comment on its safe haven properties.

The first comparison is at hand -- to gold. This has been presented in the main Results section but it needs to be stressed that gold shows favorable properties with respect to the portfolio and diversification utility compared to Bitcoin. It shows negative correlations with S\&P500 for the bulk of the distribution. The correlations grow for higher quantiles (even though they do not cross to the positive ones), i.e. the more bullish periods, which is again beneficial. And even though the correlation increases for the lowest quantiles, i.e. the most extreme negative cases, it still collapses to zero, not higher. In addition, we have the connection to VIX, where gold is again favored in most portfolio-related aspects. We see positive correlations for the distribution bulk, i.e. if uncertainty increases, the price of gold increases as well. And for the extreme cases, even though the correlation drops, it still remains positive. Therefore, even if we forget about other issues connected to Bitcoin (such as low liquidity, exchange risk, and various legal and accounting/tax issues \cite{Koutmos2018,Gandal2018,Schilling2019,Wu2019,AlYahyaee2020}), it does not outperform gold in any important aspect as a safe haven asset.

\begin{figure}[!htbp]
\begin{center}
\begin{tabular}{cc}
  \includegraphics[width=0.45\textwidth]{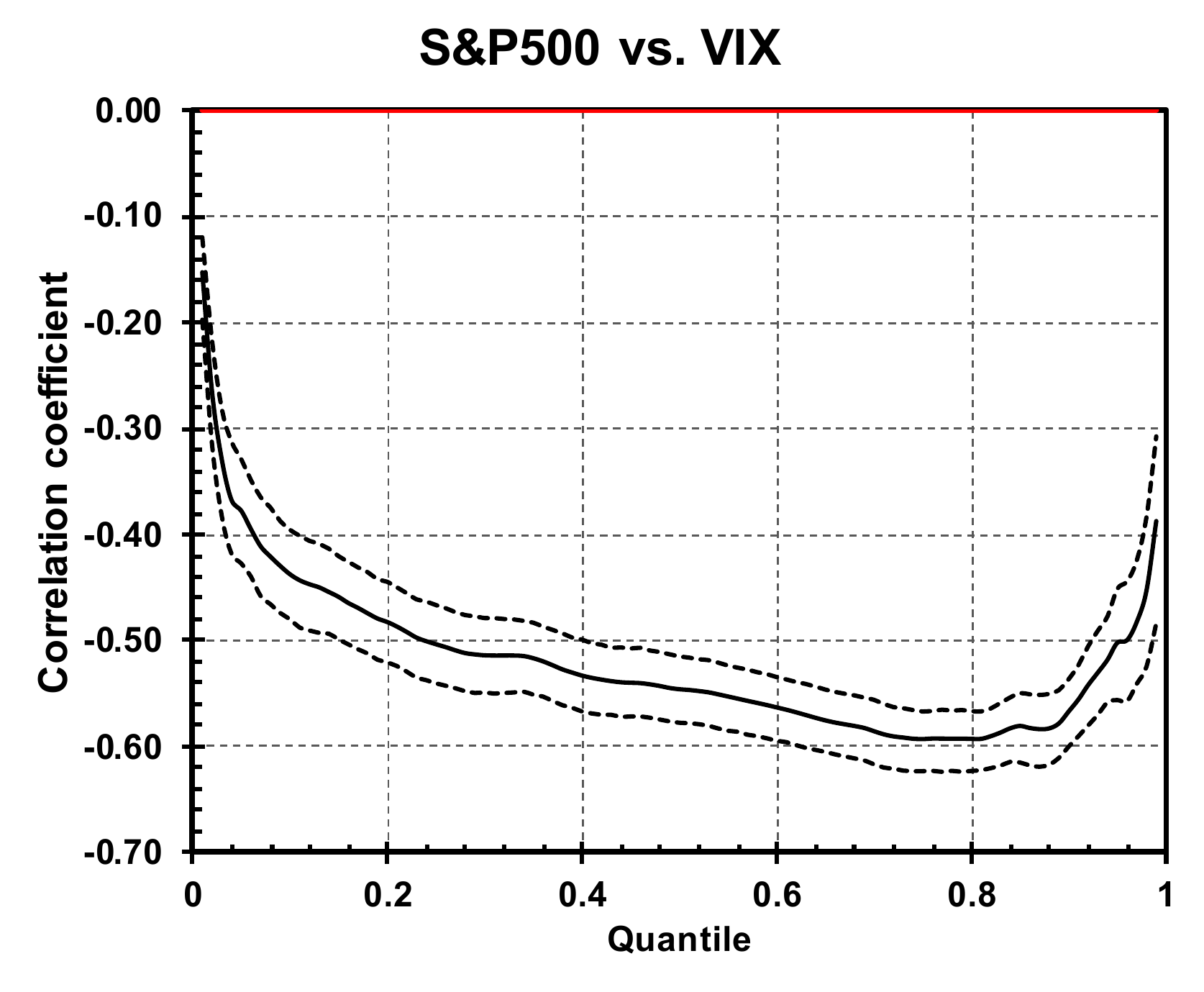}&\includegraphics[width=0.45\textwidth]{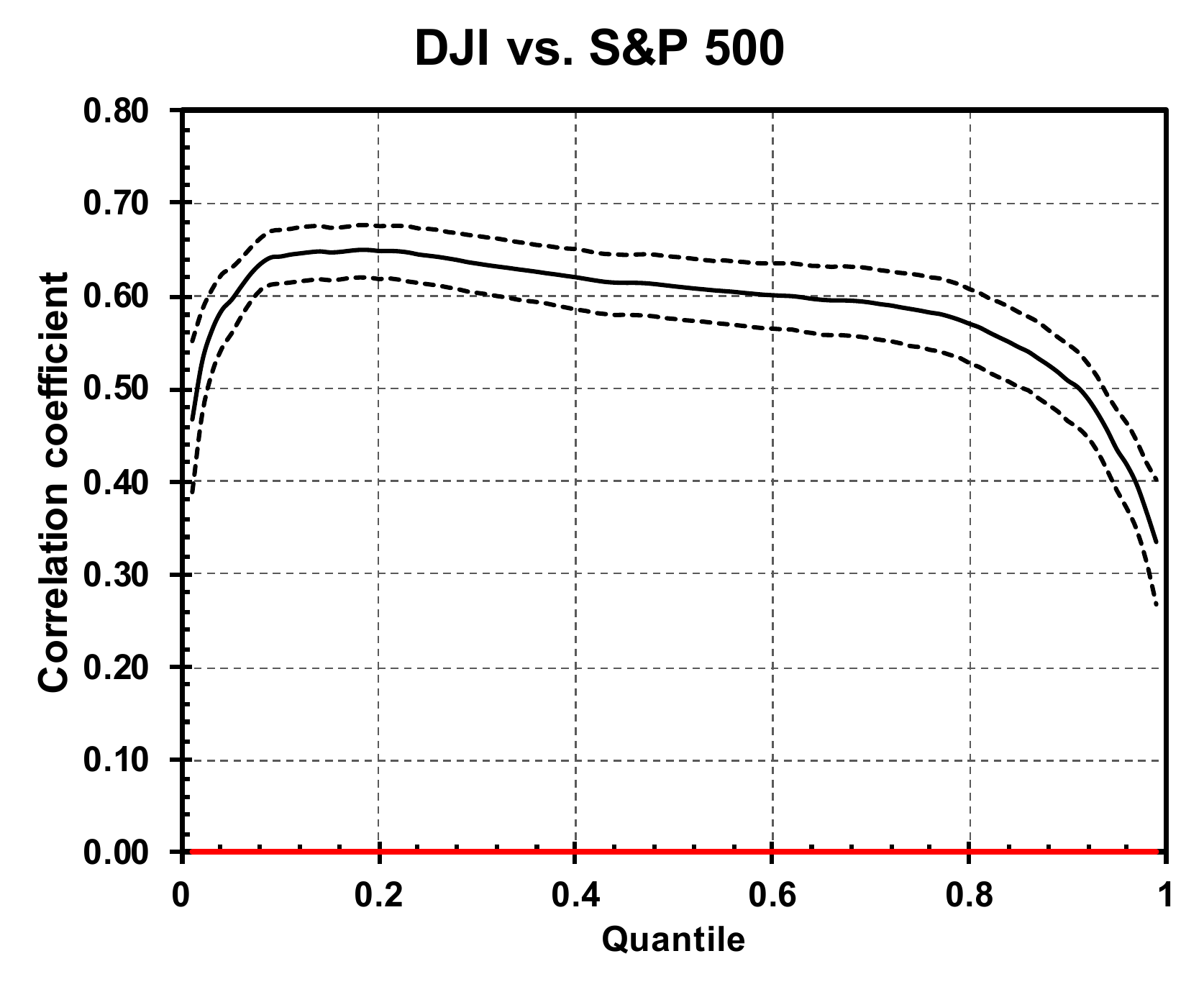}\\
    \includegraphics[width=0.45\textwidth]{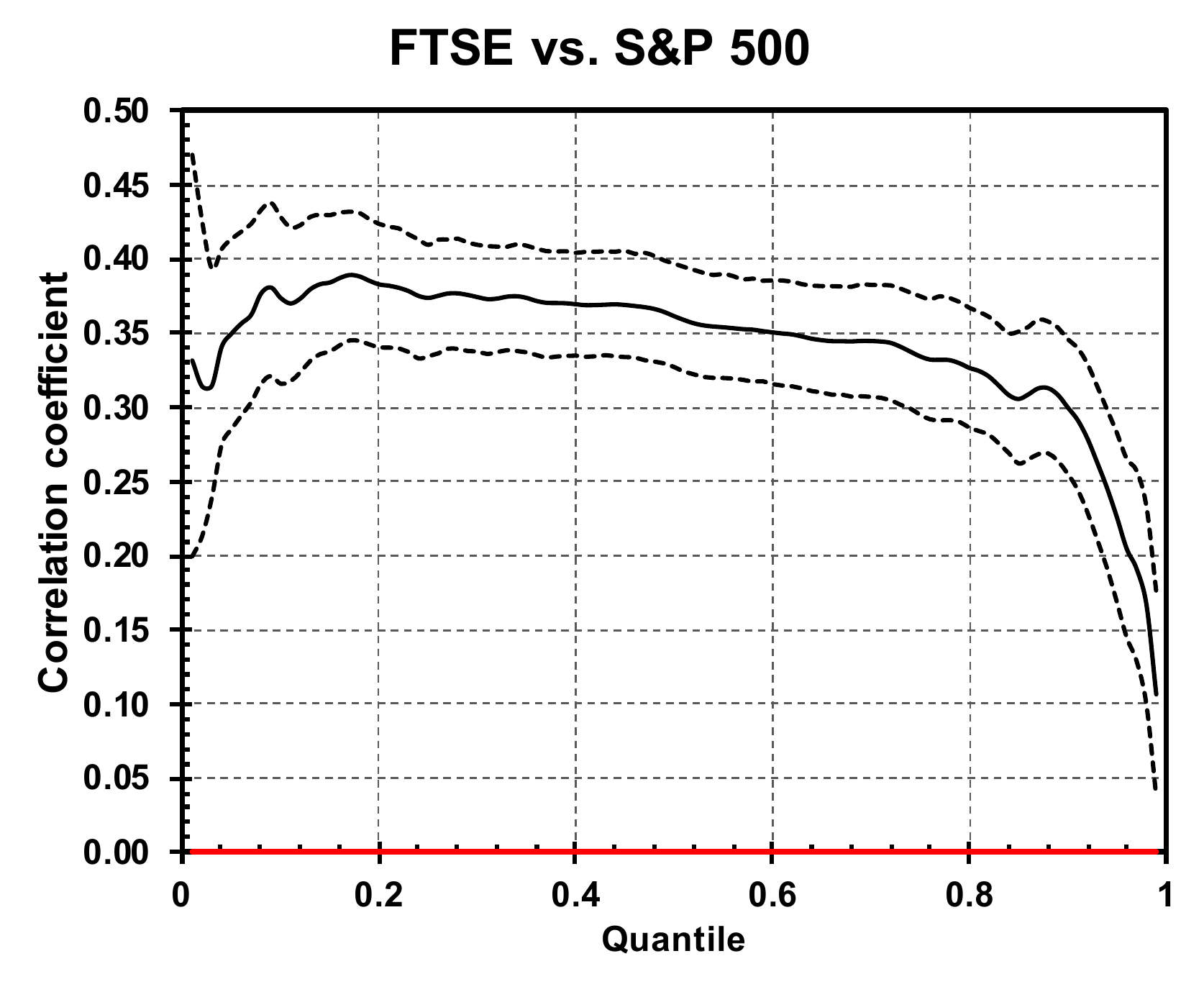}&\includegraphics[width=0.45\textwidth]{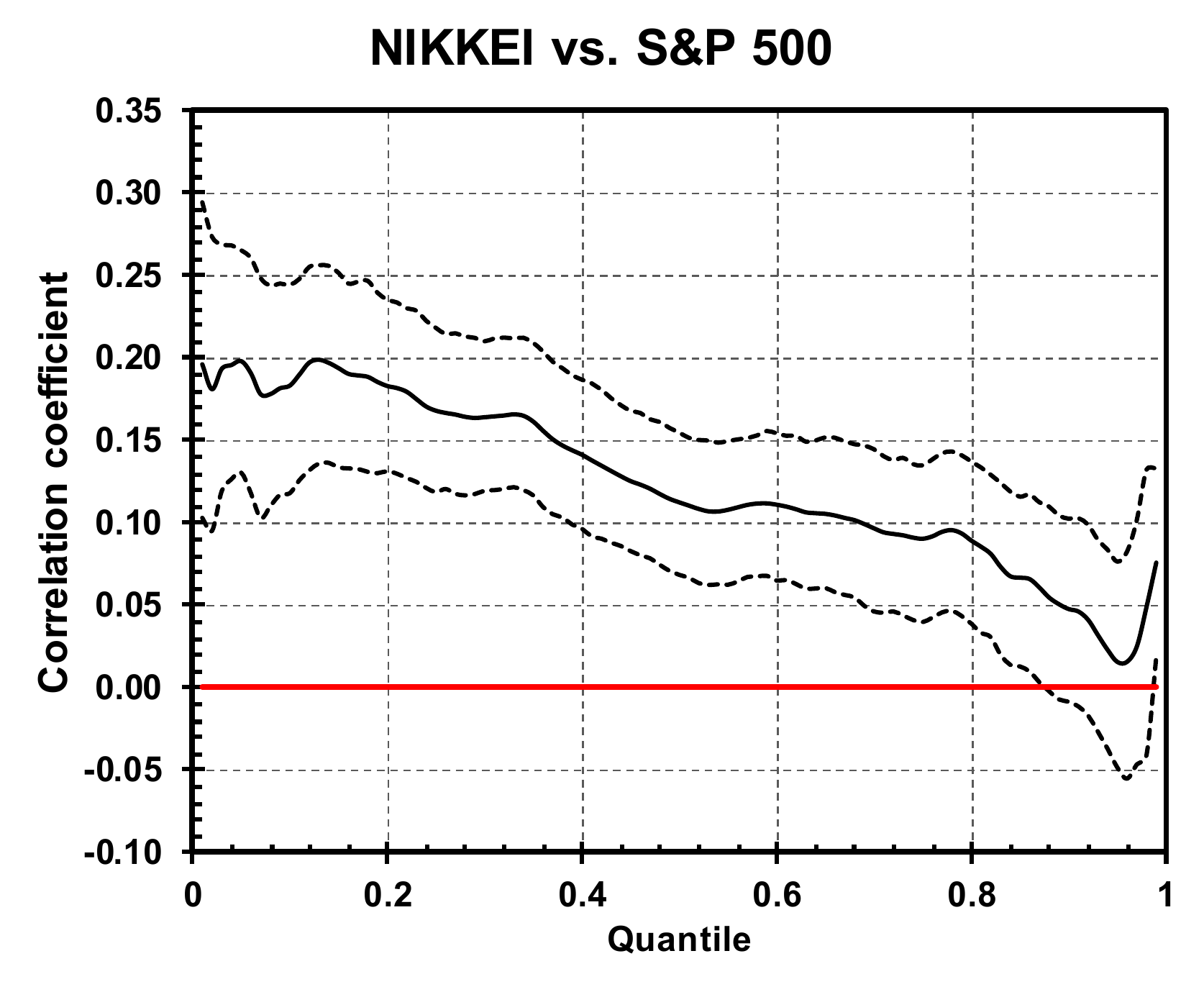}
\end{tabular}
  \caption{
Quantile correlations for S\&P500 with other assets. The notation holds from Fig \ref{BTC}.
}
  \label{Comparison}
\end{center}
\end{figure}

The second comparison is with other stock indices, mostly to get the correct grasp of the scale of correlations presented before. In Fig. \ref{Comparison}, we show the quantile correlations of S\&P500 with VIX and three other stock indices -- Dow Jones Industrial Average (DJI), Footsie 100 (FTSE) and NIKKEI 225 (NIKKEI) -- for the same period of time. There are several interesting observations. First, even for the pair of S\&P500 and DJI, the two main US stock indices (in addition to NASDAQ), the tails correlations are not as strong as one might expect -- below 0.5 for the extreme negative cases and below 0.4 for the extreme positive cases. Second, not surprisingly, S\&P500 is strongly connected to VIX. But again, its connection weakens for the extreme cases, more markedly for the calmer periods. Third, the markets are not much correlated during the extremely positive movements of the S\&P500 index where we find the quantile correlations fall to very low values for both FTSE and NIKKEI. And fourth, BTC behaves quite similarly to NIKKEI showing mild correlations for the whole spectrum of quantiles with slightly higher correlations for the extreme negative movements and practically zero correlation for the extremely positive movements. To be fair, BTC still shows more favorable low-quantile correlations than NIKKEI does, but not by much.

Overall, we show that the claim of Bitcoin being a safe haven and an alternative to gold or even being the `digital gold' are unsubstantiated and far-fetched. Although, we do not want to discredit Bitcoin in this aspect completely as the COVID-19 epidemics and the financial markets turmoil induced by it are only the first real tests to its status. Nevertheless, at this point, gold emerges as a clear winner in this safe have contest.

\section*{Acknowledgements}
Support from the Charles University PRIMUS program (project PRIMUS/19/HUM/17) and the Czech Science Foundation (project 20-17295S) is highly appreciated.


\bibliographystyle{unsrt}

\newpage

\end{document}